\documentclass[a4paper, 10pt, twocolumn]{article}

\usepackage{graphicx}
\usepackage[noblocks]{authblk}
\usepackage{mathtools}
\usepackage{algpseudocode}
\usepackage{amsmath}
\usepackage{amssymb}
\usepackage[square,sort,comma,numbers]{natbib}
\usepackage{verbatim}
\usepackage{hyperref} 
\usepackage{bm}
\usepackage{caption}
\usepackage[labelformat=empty]{caption}
\providecommand{\keywords}[1]{\noindent \textbf{Keywords:} #1}
\usepackage[usenames]{color}

\begin{document}

\title{\bf Can Mobility-on-Demand services do better after discerning reliability preferences of riders?}

\author[a]{\bf Prateek Bansal}
\author[a]{\bf Yang Liu}
\author[ ]{\bf Ricardo Daziano}
\author[ ]{\bf Samitha Samaranayake}
\affil[ ]{\bf Cornell University, Ithaca, USA}
\affil[*]{Corresponding author email: \texttt{pb422@cornell.edu}}
\affil[a]{These authors contributed equally to this work. }
\date{}
\maketitle 

\begin{abstract}
\bf \quad We formalize one aspect of \textit{reliability} in the context of Mobility-on-Demand (MoD) systems by acknowledging the uncertainty in the pick-up time of these services. This study answers two key questions: i) how the difference between the stated and actual pick-up times affect the propensity of a passenger to choose an MoD service? ii) how an MoD service provider can leverage this information to increase its ridership? We conduct a discrete choice experiment in New York  to answer the former question and adopt a micro-simulation-based optimization method to answer the latter question. In our experiments, the ridership of an MoD service could be increased by up to 10\% via displaying the predicted wait time strategically.
\\
\end{abstract}


\section{Introduction}

The usage of Mobility-on-demand (MoD) services (also known as ridehailing, ridesourcing, and transportation network companies or TNCs) is growing at a rapid rate; MoD services served 2.61 billion passengers in 2017, which is a 37\% increase from 2016~\citep{schaller2018new}. However, the systemic impact of these emerging services on the transportation ecosystem is unclear. On one hand, these services provide additional travel options to passengers and have the potential to reduce auto ownership and pollution. On the other hand, these positive effects may be offset by a potential reduction in transit share, induced travel demand due to increased convenience, and increases in vehicle miles traveled (resulting in increased traffic congestion). For instance, the results of a survey across seven cities in the United States indicate that 49\% to 61\% of the MoD served trips would not have been made at all or would have been made by transit/walking~\citep{clewlow2017disruptive}. \par

To evaluate the above-mentioned system-level impacts of MoD services, it is crucial to understand the factors affecting passengers' preferences for these services. To this end, a handful of stated preference (SP) studies have explored factors that lead to the use of MoD services over incumbent travel modes. In terms of sociodemographic characteristics, previous studies have shown that young and high-educated individuals who reside in metropolitan areas are more likely to be early adopters of these services~\citep{alemi2018influences}. To capture the effect of service levels on the mode choice, a few studies have considered traditional alternative-specific attributes such as in-vehicle travel time, trip cost, walk time, and wait time~\citep{choudhury2018modelling, konig2018analyzing, liu2018framework}. \par

Reliability of MoD services is arguably an important attribute, but it has received little attention in the literature. MoD customers experience uncertainty in both the wait time for being picked up and the travel time to the destination subsequent to the pick-up. Whereas travel time reliability on stochastic road networks has been studied in the context of vehicle routing~\citep{samaranayake2011tractable,gendreau201650th, liu2019stochastic}, the focus of this study is wait time reliability. 

When a passenger plans a trip using a TNC mobile application, the total cost and the expected time of arrival are displayed immediately. If the trip is requested, more refined wait time information is provided after the vehicle assignment. At this point, the passenger still has the freedom to cancel the trip at no cost in the next few minutes. The decision of the passenger to accept or cancel the trip generally depends on the displayed wait time and their perception of reliability, which is a manifestation of the difference between the displayed and the actual pick-up time during previous trips. Having access to rich historical data on the spatiotemporal distribution of the travel time on the network, TNCs are aware of the distribution of the time the assigned vehicle will take to get to the pick-up point and have an opportunity to increase trip acceptance rates by strategically displaying wait times. However, finding such an optimal value for a given passenger is not straightforward. More specifically, if the service provider displays a low percentile of the wait time distribution, the passenger is likely to use the service for the current trip due to the quick pick-up; However, the actual pick-up is likely to be delayed and thus, the passenger's likelihood to choose the service again in the future might be decreased due to the unreliable prediction. On the contrary, if a high percentile of the wait time distribution is displayed, the MoD service will be perceived as a reliable mode (since the vehicle is likely to pick up the passenger on time), but a higher proportion of passengers may cancel the trip due to the long wait time.

This study aims at answering the question of ``How an MoD service provider can maximize long term ride acceptance rates by strategically displaying the wait time information to users?" To the best of our knowledge, this is the first study that explores the impact of the displayed pick-up time information on the passenger's preference to use or not use the MoD service, and also explores strategies for optimizing the displayed pick-up time. \par

We take a two-pronged approach. \textit{First}, we conduct a stated preference experiment in New York City and estimate choice models to elicit how a passenger values the reliability of predicted wait time while making a choice on accepting or rejecting the MoD service. To quantify a passenger's perception of reliability, the analyst needs to record the passenger's experience during many consecutive MoD trips. To avoid fatigue in such a dynamic experiment setting, we conduct a cross-sectional choice experiment and use \textit{average pick-up delay} as a proxy for reliability. \textit{Second}, we integrate the choice model estimates in a microscopic simulator to illustrate how an MoD service provider can optimize its acceptance rate by strategically choosing the percentile of the wait time distribution to be displayed to the user. \par

The remainder of the paper is organized as follows: section \ref{sec:rel} summarizes details of the SP studies and provides key insights from the estimated choice models; section \ref{sec:micro} lays out a micro-simulation framework to optimize the display information and discusses results of the case study; conclusions and future research avenues are briefly discussed in section \ref{sec:conc}. \par

\section{Perception of Reliability}\label{sec:rel}
To understand the travel mode preferences of New Yorkers, we conducted two SP surveys in 2017. The first survey (N=978) contains a discrete choice experiment (DCE), which evaluates the impact of the wait time and reliability of MoD services on the likelihood of a passenger to use them. In the DCE, respondents were asked to choose an MoD service from a set of two services -- Service 1 is a hypothetical, fully reliable service that always picks up the passenger exactly at the displayed pick-up time, and Service 2 is a regular MoD service that can be late. Each respondent made a choice based on the displayed wait time and the average pick-up delay of these services in two choice situations. The experiment was designed by ensuring sensible trade-offs between attributes of MoD services. The final design had 24 scenarios. The attribute levels and an example of a choice situation are shown in Table \ref{tab:expdesign}. \par

\begin{table}[h!]
\centering
\caption{\footnotesize{Table 1: Discrete choice experiment (design and example)}\vspace{-.3cm}}
\label{tab:expdesign}
\resizebox{.5\textwidth}{!}{%
\begin{tabular}{lcc}
\hline
\multicolumn{3}{l}{\textbf{Attribute levels}} \\[1ex]\hline
 & \textit{Displayed Wait Time} & \textit{Average Pick-up Delay} \\ \hline
\textbf{Service 1} & \{5,10,15,20,25\} Minutes & 0 Minutes \\
\textbf{Service 2} & \{3,5,8,10,13,15,20\} Minutes & \{3,5,8,10,13\} Minutes \\ \hline
\multicolumn{3}{l}{\textbf{Example of a choice situation}} \\[1ex]\hline
 & \textit{Displayed Wait Time} & \textit{Average Pick-up Delay} \\ \hline
\textbf{Service 1} & 10 Minutes & 0 Minutes \\
\textbf{Service 2} & 8 Minutes & 5 Minutes \\ \hline
\multicolumn{3}{l}{\textbf{Note:} Both services are equivalent in terms of travel time and cost.} 
\end{tabular}}
\end{table}

\begin{table}[h!]
\centering
\caption{\footnotesize{Table 2: Binary logit estimates (reliability experiment)}\vspace{-.3cm}}
\label{tab:bl}
\resizebox{.5\textwidth}{!}{%
\begin{tabular}{lcc}
\hline
 & Estimates & t-value \\ \hline
service 2:(intercept) & -1.55 & -15.60 \\
$\log$(displayed wait time (in minutes)) & -3.88 & -15.98 \\
$\exp(\text{average pick-up delay}/\text{displayed wait time})$ & -0.78 & -15.10 \\ \hline
Loglikelihood & \multicolumn{2}{c}{-2343.6} \\
Bayesian information criterion & \multicolumn{2}{c}{4709.9} \\
McFadden R-square & \multicolumn{2}{c}{0.063} \\
Sample size (choice situations) & \multicolumn{2}{c}{978 (1956)} \\ \hline
\end{tabular}}
\end{table}

\begin{figure}[!h]
\centering
\includegraphics[width=.9\columnwidth]{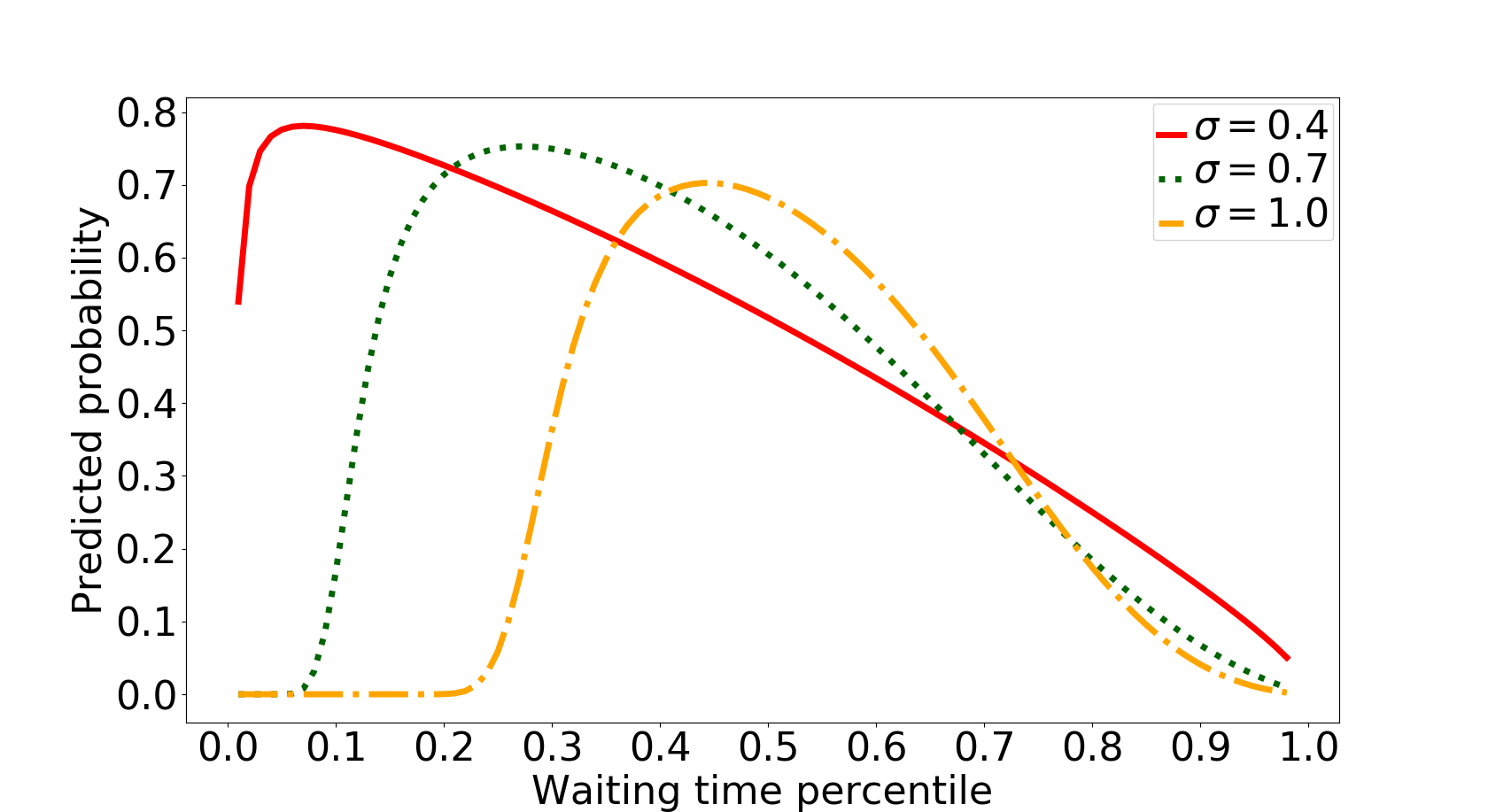}
\vspace{-.3cm}
\caption{\footnotesize{Figure 1: Predicted probability of choosing the regular MoD service.}}
\label{fig:prob_perc}
\end{figure}

We estimated binary logit models with various linear and non-linear (e.g., polynomial and logarithmic transformations) link function and chose the one with the lowest Bayesian information criterion (see Table \ref{tab:bl} for results). Since parameter estimates cannot be directly interpreted, we plot the predicted probability of using the regular MoD service at various wait time percentiles (Figure \ref{fig:prob_perc}). The assumptions and required computations to create these plots are detailed below. We assume that the wait time of the regular MoD service follows a lognormal distribution with a mean of $1.5$ minutes\footnote{The mean of the distribution is set according to the simulation using NYC taxi historical data in \citep{alonso2017demand}. } and its natural logarithm's standard deviation $\sigma$ to be $\{0.4,0.7,1.0\}$ minutes. We compute the average pick-up delay of the regular MoD service at a given percentile of the wait time distribution using $ \int_{d_w}^{\infty}f(w) \cdot (w - d_w) dw$, where $f(w)$ is a probability density function of the wait time $w$ and $d_w$ is the displayed wait time. Note that the average pick-up delay of the reliable service is zero. The wait time of the reliable service is set to 2.7 minutes in Figure \ref{fig:prob_perc}, which is the mean (1.5 minutes) plus one standard deviation in the regular MoD service wait time distribution assuming $\sigma = 0.7$. The results indicate that the optimal wait time percentile (the one corresponding to the highest choice probability) increases with the increase in $\sigma$. This trend shows a trade-off between the displayed wait time and the pick-up delay -- in cases of high uncertainty in the wait time, a higher percentile is favored since it decreases the risk of being delayed. \par
 
As emphasized earlier, the passenger's perception of reliability can be quantified by recording the experienced pick-up delays at many consecutive trips. Whereas designing an experiment to mimic such situations remains an open question, we include \textit{a segment} of such experiments in another cross-sectional SP study (N=1,512). We elicit how the pick-up delay during the most recent trip affects the New Yorkers' propensity to choose the MoD service in future trips. In this survey, respondents were given a hypothetical scenario where they choose an MoD service at a high displayed wait time (over 12 minutes) and the assigned vehicle actually gets delayed. The displayed wait time and the actual pick-up delay were varied across respondents in ranges of $[12,24]$ and $[10,15]$ minutes, respectively. Under these settings, 58.4\% respondents reported their inclination to switch to other MoD services in the future. We estimated a binary choice model to better understand the switching preferences and the results (Model 1 in Table \ref{tab:delay}) indicate an increase in this propensity with the increase in the percentage pick-up delay (pick-up delay/displayed wait time). \par 

We gathered some attitudinal insights. Around 20.8\% of respondents thought that both the MoD service provider and the driver were responsible for such pick-up delays. However, a higher proportion of service users were likely to make the driver the sole responsible for this delay vs. the MoD service (8.7\% vs. 1.7\%). This statistic indicates that designing a series of experiments can better represent reality as an MoD service actually has more opportunities to optimize the displayed wait time than those captured in experiment 1 -- the MoD service has the freedom to get delayed at these 8.7\% instances without affecting the company's reputation. \par

We further estimated ordered logit models (Models 2 and 3 in Table \ref{tab:delay}) to understand the association of such perceptions with the sociodemographic characteristics of MoD users. The parameter estimates show that Bachelor degree holders and higher income passengers are more likely to blame both the driver and the MoD service for the pick-up delay, ceteris paribus. \par 

\begin{table}[!h]
\centering
\caption{\footnotesize{Table 3: Perception of riders in case of pick-up delays} \vspace{-.3cm}}
\label{tab:delay}
\resizebox{.5\textwidth}{!}{%
\begin{tabular}{lcc}
\hline
\multicolumn{3}{l}{\textbf{Model 1: Switch to other MoD service (binary logistic regression)}} \\ \hline
\textbf{} & \textbf{Estimate} & \textbf{t-value} \\ \hline
(Intercept) & 0.43 & 5.65 \\
log(percentage difference$^{a}$) & 0.31 & 1.63 \\ \hline
Null loglikelihood & \multicolumn{2}{c}{-1026.6} \\
Model loglikelihood & \multicolumn{2}{c}{-1025.3} \\ \hline
\multicolumn{3}{l}{\textbf{Model 2: Drive is responsible (ordered logistic regression)}} \\ \hline
 & \textbf{Estimate} & \textbf{t-value} \\ \hline
percentage difference$^{a}$ & -0.73 & -3.03 \\
log(household annual income) & -0.19 & -3.07 \\
log(age) & 0.92 & 5.96 \\
bachelor degree indicator & -0.33 & -3.03\\ \hline
cutoffs & \textbf{Estimate} & \textbf{Std. error} \\ \hline
yes$\lvert$may be & 0.38 & 0.77 \\
may be$\lvert$no & 2.63 & 0.78 \\ \hline
Null loglikelihood & \multicolumn{2}{c}{-1422.8} \\
Model loglikelihood & \multicolumn{2}{c}{-1391.9} \\ \hline
\multicolumn{3}{l}{\textbf{Model 3: MoD Service is responsible (ordered logistic regression)}} \\ \hline
 & \textbf{Estimate} & \textbf{t-value} \\ \hline
percentage difference$^{a}$ & -0.44 & -1.95 \\
log(household annual income) & -0.12 & -2.14 \\
bachelor degree indicator & -0.27 & -2.64 \\
male indicator & -0.21 & -2.07 \\ \hline
cutoffs & \textbf{Estimate} & \textbf{Std. error} \\ \hline
yes$\lvert$may be & -2.88 & 0.66 \\
may be$\lvert$no & -0.88 & 0.65 \\ \hline
Null loglikelihood & \multicolumn{2}{c}{-1610.1} \\
Model loglikelihood & \multicolumn{2}{c}{-1596.9} \\ \hline
Sample size & \multicolumn{2}{c}{1512} \\ \hline
\multicolumn{3}{l}{{$^{a}$}\footnotesize{percentage difference = actual pick-up delay/displayed wait time}}
\end{tabular}}
\end{table}

In case of high displayed wait time scenarios (over 12 minutes), respondents were also asked to report the pick-up delay tolerance such that they would not have blamed any stakeholder. The results indicate that the pick-up delay tolerance of 62.5\% passengers is below 5 minutes, which implies that MoD services and drivers need to ensure punctuality to retain a good reputation among passengers if the displayed wait time is higher than 12 minutes. \par


\section{Supply-side Optimization} \label{sec:micro}
We conduct numerical experiments using taxi demand data of Manhattan, NYC to illustrate how the reliability preference estimates from the first survey can be used to optimize the ridership of an MoD service. Consistent with our mode choice model in Table~\ref{tab:bl}, taxi demand is assumed to be served by two services -- a virtual reliable service that always picks up passengers on time and a regular MoD service whose displayed wait time needs to be optimized to maximize its trip acceptance rate. \par 

We assume that the travel time between any two locations in the transportation network of Manhattan (with 4092 nodes and 9453 edges) follows a lognormal distribution \cite{liu2019stochastic}. The wait time of a passenger served by the regular MoD service is the sum of the travel time of edges on a path between the assigned vehicle's location and the passenger's origin. This is reasonable in practice because MoD service providers have access to historical data and thus can obtain the empirical distribution of the wait time for each passenger. In the absence of such information in certain situations, the provider will need to estimate the distribution. We compute the daily mean of edge travel times using the method given in \cite{santi2014quantifying}, and use it to compute the mean of the wait time distributions for each trip request. The standard deviation $\sigma$ is randomly sampled from a given range, as detailed later. \par 

Conditional on the location of the pick-up vehicle on the network, we can obtain the wait time distribution of a specific passenger under the above assumptions. Thus, the displayed wait time for a passenger can be optimized by enumerating all candidate wait time percentiles followed by computing the probability of choosing the regular MoD service for enumerated realizations of wait time using the estimated choice model in Section~\ref{sec:rel}.\footnote{We resort to the enumeration-based optimization because the derivative of the choice probability is analytically intractable due to a complex integral in the average pick-up delay expression.}  \par

The trip acceptance rate of the regular MoD service is the mean probability of choosing the service by each passenger, but this probability depends on the location of the assigned vehicle at the time of the trip request. However, in the simulation, vehicle locations are dependent on the complex supply-demand dynamics. To handle this concern, we modify the objective to the expected probability of choosing the regular MoD service, where the expectation is taken over the mean travel time of the paths that vehicles use to pick-up the passenger at the origin node (O-node, henceforth). Based on our earlier assumption, this mean travel time is the same as the \textit{mean wait time}. Specifically, we adopt a two-step strategy to compute the above-mentioned expected probability. In the first step, we run the simulation using a week's demand data. We track the vehicles serving the trip requests at each O-node and record the mean wait time. By computing the proportion of vehicles with each mean wait time value, we obtain a probability mass function of mean wait time for each O-node. The probability mass function reflects the likelihood of having a vehicle on the network that can serve the demand at each O-node with a certain mean wait time. In the second step, conditional on the mean wait time (i.e. the location of the pick-up vehicle), we generate a wait time distribution and find the optimal wait time percentile to display for passengers at each O-node which maximizes the conditional probability of choosing the regular MoD service. Then, we compute the expected probability of a passenger at each O-node to choose the regular MoD service, where the expectation is taken over the mean wait time with the probability mass function obtained in Step 1. The details of these steps are provided below. \par

\textbf{Step 1:} We run the MoD service simulator for taxi demand of the first week of May 2013. In this simulation, we first predict the probability of each passenger choosing the regular MoD service considering the mean wait time as the displayed wait time and assuming no pick-up delays. Then we take a draw from a random variable which follows a uniform distribution between 0 and 1. The passenger chooses the regular MoD service if the generated random number is below its predicted probability. If the passenger ends up choosing the regular MoD service, the passenger is served within the simulator; otherwise, no further action is required. \par

Over the period of a week, consider that the trip requests at node $i$ are served by vehicles with a set of mean wait time $t \in T_i$. By computing proportion of each mean wait time value $t$, we obtain the probability mass $f_{i}(t)$.\footnote{We are making an implicit assumption that the probability mass $f_{i}$ does not change with the displayed wait time.}\par

\textbf{Step 2:} For each mean wait time $t$, the optimal wait time value to display for passengers at node $i$ can be obtained by enumerating the candidate percentiles in the wait time distribution and finding the one which maximizes: \par

\begin{equation}
    P_{\text{MoD}}^{i, t, *} = \max\limits_{d_w \in \mathcal{W}} P_{\text{MoD}}^{i, t} (d_w)
\end{equation}

where $\mathcal{W}$ is a set of the candidate wait times to be displayed, $P_{\text{MoD}}^{i, t} (d_w)$ is the predicted probability of choosing the regular MoD service by the demand at node $i$ given the mean wait time $t \in T_i$ and the displayed wait time $d_w$. \par

Given the probability mass function $f_{i}$ of the mean wait time, we can compute the expected probability $\mathbb{E}_{i}$ of choosing the regular MoD service by passengers at node $i$ under the optimal displaying strategy:
\begin{equation}
    \mathbb{E}_{i}  = \sum\limits_{t \in T_{i}} f_{i}(t) \cdot P_{\text{MoD}}^{i, t, *}
\end{equation}
The expected trip acceptance rate of MoD is then obtained by taking the weighted average of $\mathbb{E}_{i}$ across nodes, where weight of node $i$ is the demand at that node. \par

We consider a fixed fleet size of 1500 single-occupancy vehicles for the regular MoD service and use the state-of-the-art MoD simulator~\citep{alonso2017demand} to assign vehicles to passengers. When the taxi demand is served by a fleet at such scale, the average of the passengers' wait time lies in the range of 1-2 minutes~\citep{alonso2017demand}, and we set the wait time of the reliable service to 2 minutes. For each request, $\sigma$ is randomly chosen from a predefined range and used to generate the wait time distribution. Then, we enumerate the percentile in the range $[20, 80]$ with a step of 1 percentile in the wait time distribution. To capture the effect of the extent of uncertainty in the wait time, we conduct experiments under three different $\sigma$ ranges: $[0.1, 1.0]$, $[0.4, 1.0]$, and $[0.8, 1.0]$ minutes. We report the average of the optimal percentiles across the entire demand. We also compute the trip acceptance rate of the regular MoD service when i) the optimal percentile of the wait time distribution and ii) the expected value of the wait time is displayed to passengers. The results are summarized in Table \ref{tab:share}. Similar to Figure \ref{fig:prob_perc}, the optimal wait time percentile increases with the increase in the distribution's variance to circumvent the high risk of being late. By optimizing the displayed wait time information, the regular MoD service could gain an additional 9.9\% to 11.7\% trip acceptance rate, which is about 1,207,665 to 1,421,247 more trips in a month.\par

\begin{table}[!h]
\centering
\caption{\footnotesize{Table 4: Comparison of performance metrics} \vspace{-.3cm}}
\label{tab:share}
\resizebox{.5\textwidth}{!}{%
\begin{tabular}{ccccc}
\hline
\begin{tabular}[c]{@{}c@{}} Std. dev. \\ ($\sigma$) range \end{tabular} & \begin{tabular}[c]{@{}c@{}}Trip acceptance rate \\ using EWT{$^{b}$} \end{tabular} & \begin{tabular}[c]{@{}c@{}}Trip acceptance rate  \\ using optimal DWT{$^{c}$} \end{tabular} & \begin{tabular}[c]{@{}c@{}} Increase in \\ Trip acceptance rate \end{tabular} & \begin{tabular}[c]{@{}c@{}}Optimal wait \\ time percentile \end{tabular} \\ \hline
$[0.1, 1.0]$ & 0.338 & 0.437 & 0.099 & 0.258 \\
$[0.4, 1.0]$ & 0.334 & 0.451 & 0.117 & 0.288 \\
$[0.8, 1.0]$ & 0.328 & 0.441 & 0.113 & 0.391 \\ \hline
\multicolumn{5}{l}{{$^{b}$}\footnotesize{EWT: expected wait time}; {$^{c}$}\footnotesize{DWT: displayed wait time}}
\end{tabular}%
}
\end{table} 

\section{Conclusions} \label{sec:conc}
We provide a framework that integrates the demand and supply sides to understand the impacts of pick-up time reliability in the context of MoD services. On the demand side, we estimate the impact of pick-up delays of MoD services on the likelihood of passengers to choose these services. On the supply side, we incorporate the demand-side estimates in the micro-simulation-based optimization to find the optimal wait time that MoD services should display to attract more passengers. In numerical experiments, we illustrate that an MoD service can gain an additional around 10\% trip acceptance rate by just displaying appropriate wait time information. \par

This study, however, relies on  a simplistic experiment design and various assumptions on the wait time distribution. Therefore, two key research avenues emerge. First, a more comprehensive series of discrete choice experiment can be designed to explain the effect of the displayed information on the passenger's preference to use the MoD services in their current trip and in future trips. Second, future research in collaboration with TNCs can help in adding more realism to this study by calibrating the proposed optimization method with the real distribution of the wait time.  \par

\bibliography{short_version}
\bibliographystyle{unsrt}
\end{document}